\documentclass[fleqn,10pt]{wlscirep}
\usepackage[utf8]{inputenc}
\usepackage[T1]{fontenc}
\usepackage{booktabs}

\title{Asymmetric Little-Parks Oscillations in Full Shell Double Nanowires}

\author[1,2]{Alexandros Vekris}
\author[1]{Juan Carlos Estrada Salda\~{n}a}
\author[3]{Joeri de Bruijckere}
\author[1]{Sara Lori\'{c}}
\author[1]{Thomas Kanne}
\author[1]{Mikelis Marnauza}
\author[1]{Dags Olsteins}
\author[1]{Jesper Nyg{\aa}rd}
\author[1]{Kasper Grove-Rasmussen}
\affil[1]{Center for Quantum Devices, Niels Bohr Institute, University of Copenhagen, 2100 Copenhagen, Denmark}
\affil[2]{Sino-Danish College (SDC), University of Chinese Academy of Sciences}
\affil[3]{Kavli Institute of Nanoscience, Delft University of Technology, 2628 CJ Delft, The Netherlands}

\begin{abstract}
Little-Parks oscillations of a hollow superconducting cylinder are of interest for flux-driven topological superconductivity in single Rashba nanowires. The oscillations are typically symmetric in the orientation of the applied magnetic flux.
Using double InAs nanowires coated by an epitaxial superconducting Al shell which, despite the non-centro-symmetric geometry, behaves effectively as one hollow cylinder, we demonstrate that a small misalignment of the applied parallel field with respect to the axis of the nanowires can produce field-asymmetric Little-Parks oscillations. These are revealed by the simultaneous application of a magnetic field perpendicular to the misaligned parallel field direction. The asymmetry occurs in both the destructive regime, in which superconductivity is destroyed for half-integer quanta of flux through the shell, and in the non-destructive regime, where superconductivity is depressed but not fully destroyed at these flux values.

\end{abstract}
\begin{document}

\flushbottom
\maketitle

\thispagestyle{empty}

\section*{Introduction}

The recent observation of signatures of flux-induced topological superconductivity in individual semiconductor nanowires coated by a shell of superconducting Al has brought the Little-Parks (LP) effect into the spotlight~\cite{little1962observation,Vaitiekenas2020Mar}. The reduced dimensions of this core-shell system make possible interesting manifestations of this effect~\cite{vaitiekenas2020anomalous}. The thinness of the shell results in fluxoid, rather than flux quantization~\cite{deaver1961experimental,doll1961experimental}. Depending on the ratio between the coherence length, $\xi$, and the diameter of the nanowire, $d$, which determines the diameter of the shell, the LP  oscillations can either exhibit a reduced critical temperature, $T_\mathrm{c}$, at half integer values of flux quantum $\Phi_0$ (non-destructive regime, $\xi \ll d$), or $T_\mathrm{c}$=0 (destructive regime, $\xi \gg d$). In the destructive regime \cite{liu2001destruction}, the application of a magnetic field perpendicular to the nanowire simultaneously with a field which threads magnetic flux through the shell can provoke the emergence of an anomalous metallic phase between nearby LP domes~\cite{vaitiekenas2020anomalous}.

While the use of single nanowires for investigations of the LP effect is at an early stage~\cite{Vaitiekenas2020Mar,vaitiekenas2020anomalous,Valentini2020Aug,Sabonis2020Oct,razmadze2020quantum}, the use of double nanowires is still unheard of. Double nanowires covered by a half/full superconducting shell are of interest for exploring robust manifestations of topological superconductivity, such as Majorana zero modes,  the topological Kondo effect and parafermionic modes~\cite{Beri2012Oct,Klinovaja2014Jul}. The realization of the two former could benefit from the advantages of the potentially vortex-induced topological superconductivity investigated in single-nanowire devices due to the LP effect~\cite{Vaitiekenas2020Mar,PenarandaPhysRevRes2020}.

In hollow cylinders made of thin superconductor materials, including the nanowire shells described above, the applied magnetic field, $B$, needs to be properly aligned with the axis of the cylinder so as to maximize the critical field, $B_\mathrm{c}$, at which the LP oscillations die out due to bulk destruction of superconductivity. This can be done by mechanical alignment of the sample to the axis of an external coil, or by field rotation using two-axis or three-axis vector coils to align the field with the sample orientation. Both of these ways of alignment are subject to error due to finite experimental resolution.

Here, we report the Little-Parks effect in closely-spaced InAs double nanowires fully covered by a thin epitaxial superconducting Al shell\cite{Kannepreprint2021}. The nanowires are used as a template to shape the shell. Therefore, while the shell could potentially behave as two connected but individual hollow superconducting cylinders, we find in this work, by comparing our measurements to a mean-field model, that the shell actually behaves as a \textit{single} cylinder. 
In addition to demonstrating the single-cylinder behavior of the shell of the nanowires, we show a way of breaking the $T_\mathrm{c}(B)=T_\mathrm{c}(-B)$ symmetry of the LP oscillations, which relies on $B$ misalignment. As the single-cylinder model predicts the presence of the asymmetry for any misalignment, the degree of asymmetry can be used as an accurate measurement of the degree of misalignment of the field with the long axis of the sample. 
For completeness, we note that similar double nanowires, however, with only half-shell superconductor coverage are addressed in several parallel works \cite{Kannepreprint2021,Kurtossypreprint2021,Vekrispreprint2021}.


\section*{Results}

\subsection*{Setup}

\begin{figure}[h!]
    \centering
    \includegraphics[width=1\textwidth]{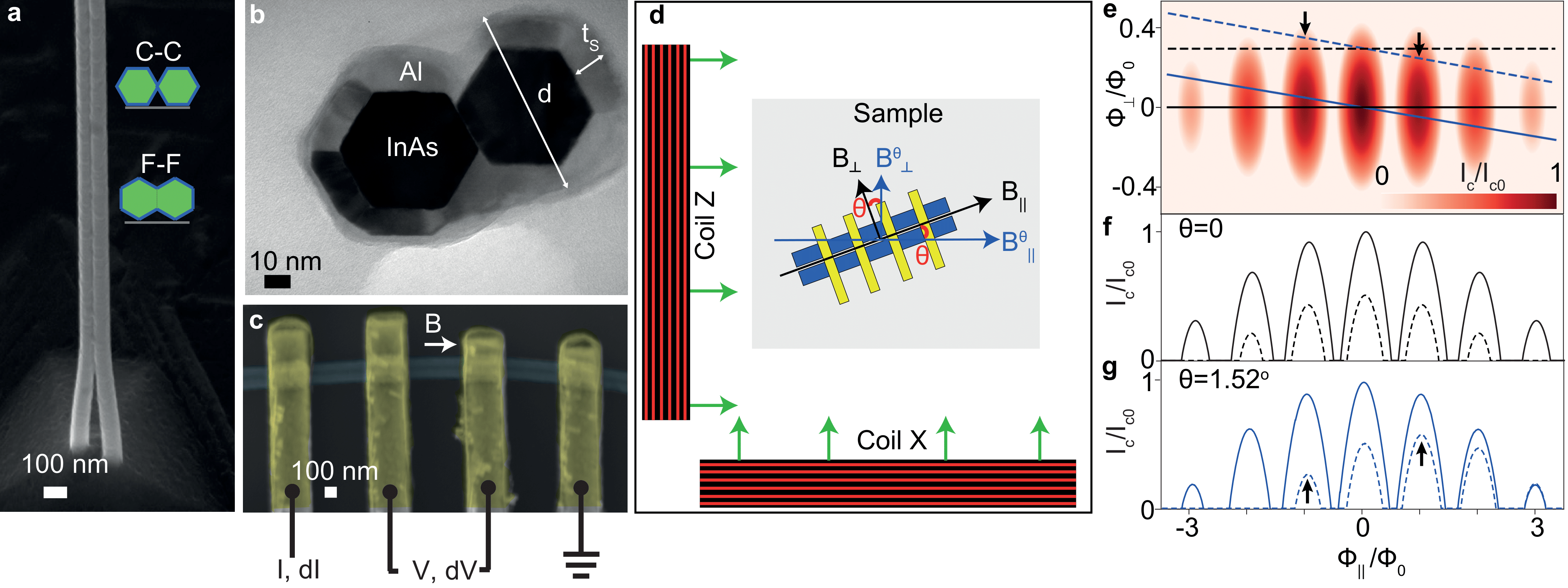}
    \caption{\textbf{Experimental setup and model for asymmetric Little-Parks effect.} \textbf{(a)} Scanning electron micrograph of as-grown InAs double nanowire covered with in-situ deposited epitaxial Al. The inset shows two possible relative configurations of the six-faceted nanowires. \textbf{(b)} Transmission electron micrograph of a $\approx80$ nm-thick slice of double nanowires. \textbf{(c)} False-colored scanning electron micrograph of a device. \textbf{(d)} Schematics of the magnetic field setup. \textbf{(e-g)} $I_\mathrm{c}$ dependence on parallel and perpendicular magnetic fluxes calculated with a single hollow superconducting cylinder model, showing an asymmetric Little-Parks effect due to field misalignment in the case of the dashed blue curve. See text for details.}
    \label{device}
\end{figure}

The InAs double nanowires are grown by the vapor-liquid-solid method, with Au droplets as growth catalysts. The growth is followed by in-situ Al epitaxy~\cite{krogstrup2015epitaxy,Kannepreprint2021}. A typical example of the as-grown Al-coated double nanowires is shown in the scanning electron micrograph Fig.~\ref{device}a. 
Despite being grown from gold droplets which are separated by $>100$ nm, the nanowires usually clamp together at their upper segments. The clamped part constitutes the bulk of the double nanowires and it is the part investigated in this work. Fig.~\ref{device}b shows a transmission electron micrograph of a thin cross-sectional slice of the clamped part of a double nanowire. The two nanowires (in black) have an hexagonal cross section with six facets each. They are covered by Al (in grey) on their five exterior facets. Their remaining facets face each other with a small relative misalignment. There is no substantial Al in between. The inset schematics in Fig.~\ref{device}a show the possible relative orientations of the nanowires: 1) facet-to-facet (F-F), as in Fig.~\ref{device}b, and 2) corner-to-corner (C-C). The relative orientations are chosen by properly designing the positions of the gold droplets through electron beam lithography; however, the exact relative positions are subject to variability\cite{Kannepreprint2021}. The primary sources of misalignment may relate to the Au particle formation mechanism and to Au particle diffusion. Slices such as the one in Fig.~\ref{device}b taken from other double nanowires show different relative placement and distances between the nanowires, reflecting this variability and the possibility that the nanowires do not fully clamp before the Al is deposited. Both C-C and F-F devices were investigated, with no significant differences found in most devices.


To characterize the Little-Parks effect in the superconducting Al shell of the double nanowires, we performed four-terminal differential resistance, $dV/dI$, measurements in current-biased mode in devices with the layout of the one shown in the scanning electron micrograph of Fig.~\ref{device}c. The measurements were done in a dilution refrigerator with a base temperature of $\mathrm{T=30\,mK}$. In the devices, the Al shell was contacted with Ti/Au leads following milling of the native Al oxide. To record $dV/dI$, a device was biased with a small lock-in excitation $dI=10$ nA superposed to a DC current $I$, and the ensuing AC and DC voltage drops, $dV$ and $V$, were measured with a lock-in amplifier technique and a digital multimeter, respectively.

Using a two-axis vector magnet, we apply on the sample an external magnetic field, $\mathbf{B}$, which can be divided into parallel, $B_\parallel$, and perpendicular, $B_\perp$, components to the axis of the double nanowires. $B_\parallel$ is used to thread flux through the shell of the nanowires for the LP effect and to eventually fully destroy superconductivity at $B_{c\parallel}$, the parallel critical field of the shell, while the only role of $B_\perp$ is to suppress superconductivity until full destruction at $B_{c\perp} \ll B_{c\parallel}$. $\mathbf{B}$ is nominally applied in the plane of the sample; a small out-of-plane misalignment should not alter qualitatively the conclusions presented here. The setup is schematically shown in Fig.~\ref{device}d. Nominally, $B_\parallel$ is perfectly aligned to the long axis of the sample, while $B_\perp$ is orthogonal to this direction. These two directions are represented by black arrows in Fig.~\ref{device}d. We denote as $B^\theta_\parallel$ and $B^\theta_\perp$ the two components of $\mathbf{B}$ which are instead misaligned by an angle $\theta$ from $B_\parallel$ and $B_\perp$, respectively. The effect of such misalignment is systematically studied.

\subsection*{Single-cylinder model and expected asymmetries in Little-Parks oscillations}

Little-Parks oscillations of $T_\mathrm{c}$ are expected to follow a $T_\mathrm{c}(B)=T_\mathrm{c}(-B)$ symmetry. This symmetry can be exceptionally broken in the vicinity of a hysteretic ferromagnet~\cite{Golubovic2003Nov,Samokhvalov2009May}. Here, we discuss instead an intrinsic asymmetry of LP oscillations due to minor field misalignment that may occur in experiments.
To show the expected effect of the misalignment angle $\theta$ on the LP oscillations, we employ the hollow thin-walled superconducting cylinder model used before in Ref.~\cite{vaitiekenas2020anomalous} to fit LP data in single InAs nanowires coated by an Al shell\cite{groff1968fluxoid,schwiete2009persistent,Schwiete2010Dec}. In this model, $T_\mathrm{c} (\mathbf{B})$ is provided by

\begin{equation}
\ln\Bigg(\frac{T_c(\alpha)}{T_\mathrm{c0}}\Bigg)=\Psi\Bigg(\frac{1}{2}\Bigg)-\Psi\Bigg(\frac{1}{2}+\frac{\alpha}{2\pi T_c(\alpha)}\Bigg)    \label{eq:2}
\end{equation}

 where $\Psi$ is the Digamma function~\cite{abrikosov1960contribution} and $T_\mathrm{c0}=T_\mathrm{c} (\mathbf{B}=0)$. The Cooper-pair breaking parameter~\cite{schwiete2009persistent,sternfeld2011magnetoresistance,dao2009destruction}, $\alpha=\alpha_\parallel(B_\parallel)+\alpha_\perp(B_\perp)$, contains the effects of both $B_\parallel$ and $B_\perp$ on $T_\mathrm{c}$~\cite{tinkham2004introduction,rogachev2005influence}:

\begin{equation}
    \alpha_\parallel=\frac{4\xi^2 T_\mathrm{c0}}{A_\parallel}\Bigg[\Bigg(n-\frac{\Phi_\parallel}{\Phi_0}\Bigg)^2 + \frac{t_\mathrm{s}^2}{d_F^2} \Bigg(\frac{\Phi_\parallel^2}{\Phi_0^2} + \frac{n^2}{3}\Bigg)\Bigg], \\\
    \alpha_\perp=\frac{4\xi^2 T_\mathrm{c0}}{A_\perp}\frac{\Phi_\perp^2}{\Phi_0^2} 
    \label{eq:1}
\end{equation}

The LP oscillations are encoded in $\alpha_\parallel(B_\parallel)$ given in Eq.~\ref{eq:1}, where $\xi$ is the coherence length, $d_F$ is the diameter of the cylinder, $t_\mathrm{S}$ is its wall thickness, $\Phi_\parallel=B_\parallel A_\parallel$ is the magnetic flux threading the cylinder of cross section $A_\parallel=\frac{\pi}{4} d_F^2$, and $n$ is the number of flux quanta threaded through the cylinder. The first term in $\alpha_\parallel(B_\parallel)$ oscillates with $\Phi_\parallel$ and attains a maximum for half-integer $\frac{\Phi_\parallel}{\Phi_0}$, while it is zero for integer values of this ratio. In ultra thin-walled cylinders (i.e., $t_\mathrm{s}/d_F \ll 1$), it dominates over the second term. If the $t_\mathrm{s}/d_F$ ratio cannot be neglected, as it is the case in our devices, then the second term provokes small shifts of the LP $T_\mathrm{c}$ maxima.
In turn, the Cooper-pair breaking effect of $B_\perp$ is given by $\alpha_\perp(B_\perp)$ in Eq.~\ref{eq:1}, where $\Phi_\perp=B_\perp A_\perp$, and  $A_\perp$ \cite{PhysRevB.76.094511} is a free fitting parameter. 


To convert the misaligned fields $B^\theta_\parallel$ and $B^\theta_\perp$ shown in the scheme in Fig.~\ref{device}d into $B_\parallel$ and $B_\perp$, we use:

\begin{equation}
    B_\parallel=\big[B_\parallel^\theta cos(\theta)-B_\perp^\theta sin(\theta)\big], \\\ B_\perp=\big[B_\parallel^\theta sin(\theta)+B_\perp^\theta cos(\theta)\big]  \label{eq:4}
\end{equation}

The critical current, $I_\mathrm{c}$, which is the main quantity which we measure in our devices, is given by $T_\mathrm{c}$:

\begin{equation}
I_\mathrm{c}(\alpha)=I_\mathrm{c0}\Bigg(\frac{T_\mathrm{c}(\alpha)}{T_\mathrm{c0}}\Bigg)^{3/2}
\end{equation}

where $I_\mathrm{c0}=I_\mathrm{c} (\mathbf{B}=0)$.

In Fig.~\ref{device}e, we show a calculated colormap of $I_\mathrm{c}$ versus $\Phi_\parallel$ and $\Phi_\perp$. The colormap shows oscillations of the magnitude of $I_\mathrm{c}$ against $\Phi_\parallel$, and a monotonic $I_\mathrm{c}$ reduction against $\Phi_\perp$. The oscillations come as a direct consequence of the LP oscillations of $T_\mathrm{c}$. 

Lines in Fig.~\ref{device}e indicate four types of $\mathbf{B}$ trajectories provided by the vectorial combination of $B_\parallel$ and $B_\perp$. In Fig.~\ref{device}f, we show the $I_\mathrm{c}$ dependence in trajectories for $\theta=0$, i.e. zero field misalignment. These trajectories either cross the origin in Fig.~\ref{device}e, as in the case of the solid black line ($B_\perp=0$), or are parallel to the horizontal axis, as in the case of the dashed black line ($B_\perp>0$). The corresponding oscillations of $I_\mathrm{c}$ are perfectly $\pm \Phi_\parallel$-symmetric.

The behavior of the $I_\mathrm{c}$ LP oscillations against $\Phi_\parallel$ is different when $\theta>0$, i.e., for finite field misalignment. Fig.~\ref{device}g shows the case when $\theta=1.52^\circ$. Whereas the tilted trajectory which crosses the origin in Fig.~\ref{device}e, given by the solid blue line, still gives rise to perfectly $\pm \Phi_\parallel$-symmetric $I_\mathrm{c}$ oscillations in Fig.~\ref{device}g, the tilted trajectory given by the dashed blue line which is shifted vertically by $B^\theta_\perp>0$ results in strongly \textit{asymmetric} LP oscillations. Black arrows in Figs.~\ref{device}e,g point to asymmetries in the height of the first LP lobes best seen in the dashed blue curve in Fig.~\ref{device}g. In the same curve, due to misalignment, the second and third lobes at negative $\Phi_\parallel$ are absent. A secondary consequence of finite misalignment is that, even for $B^\theta_\perp=0$, the magnitude of the LP lobes away from $\Phi_\parallel=0$ is always smaller than for perfect alignment; e.g., compare the solid blue curve in Fig.~\ref{device}g with the solid black curve in Fig.~\ref{device}f.

\begin{figure}[h!]
    \centering
    \includegraphics[width=1\textwidth]{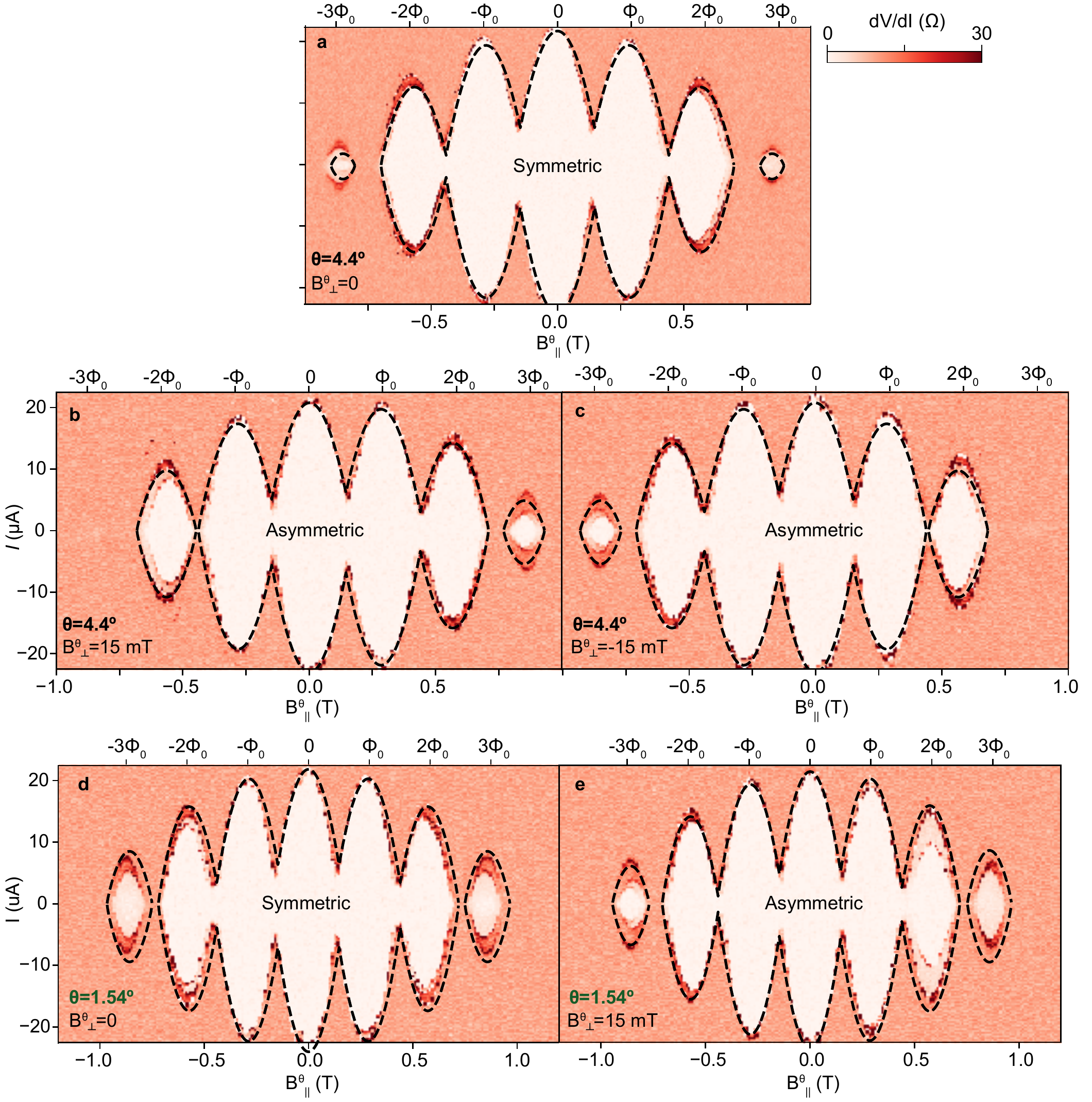}
    \caption{\textbf{Evidence for single hollow superconducting cylinder behavior and asymmetric Little-Parks effect in the non-destructive regime (Device 1).} \textbf{(a-e)} Colormaps of differential resistance, $dV/dI$, versus DC bias current, $I$, and magnetic field approximately parallel to the axis of the double nanowires, $B^\theta_\parallel$, with a small misalignment angle $\theta$ indicated in each plot. A small, constant magnetic field approximately perpendicular to the axis of the double nanowires, $B^\theta_\perp$, misaligned by the same angle $\theta$, was also applied in (b,c,e), as indicated in each of these plots. Dashed lines in (a-e) are calculations of critical current, $I_\mathrm{c}$, using the single hollow superconducting cylinder model. Fit parameters are given in Tab.~\ref{Tab1}.}
    \label{fig2}
\end{figure}

\subsection*{Single-cylinder behavior in double nanowires}

Before discussing asymmetries in the Little-Parks oscillations in the measured data, we first demonstrate the single hollow superconducting cylinder behavior of the superconducting shell of the investigated double-nanowire devices by fitting our experimental results to the above model. To do this, we focus on the well-established symmetric LP effect at $B_\perp=0$ in a device in the non-destructive regime.

In the colormap of Fig.~\ref{fig2}a, $dV/dI$ is plotted as a function of $B^\theta_\parallel$ and $I$, for $B_\perp=0$. The boundary of the white-color lobes, inside of which $dV/dI=0$ or $dV/dI \approx 0$ and outside of which $dV/dI=R_\mathrm{N}$, the normal-state resistance, corresponds to LP oscillations of the critical current, $I_\mathrm{c}$. These are dependent on $T_\mathrm{c}$. Data showing oscillations of $T_\mathrm{c}$ is shown in the Supplementary Materials (SM). The $I_\mathrm{c}$ oscillations are nearly symmetric in $B^\theta_\parallel$, with small asymmetries related to a finite small ($<20$ mT) remanence in the X and Z coils of the vector magnet (shown in SM).

By fitting the measured $I_\mathrm{c}$ to the corresponding values calculated by our model (dashed lines in Fig.~\ref{fig2}a), we obtain $\theta=1.54^\circ$. The good quality of the fit indicates that the superconducting Al shell of the two nanowires can be faithfully described as a single shell, despite the ellipsoidal cross section \cite{Daumens1998Nov}. To produce the fit, we equate $d_F$ with an effective cylinder diameter $d^*$, which corresponds to the diameter of a circle with the same area as the cross section of the two nanowires. Fit parameters are provided in Tab.~\ref{Tab1}. In the single cylinder model, the ratio $d^*/\xi$ determines whether destructive (for $d^*/\xi<1.2$, $T_\mathrm{c}=0$ at $\frac{\Phi_\parallel}{\Phi_0}=n/2$) or non-destructive (for $d^*/\xi>1.2$, $T_\mathrm{c}>0$ at $\frac{\Phi_\parallel}{\Phi_0}=n/2$) regimes arise~\cite{schwiete2009persistent}. In the SM, we compile the $d^*/\xi$ ratio obtained from fits with our model in five different double-nanowire devices, and show that this prediction holds also well in our devices. The quality of these fits indicates that these five devices behave as single hollow superconducting cylinders.

\begin{table}[]
\caption{Model parameters used to fit data of Devices 1 and 2 in Figs.~\ref{fig2} and \ref{fig3}. From left to right, coherence length ($\xi$), effective perpendicular flux area ($A_\perp$), effective parallel flux area ($A_\parallel$), and ratio of shell thickness ($t_\mathrm{s}$) to effective single cylinder diameter ($d^*$). For parameter extraction methods, see Methods.}
\label{Tab1}
\begin{tabular}{@{}ccccc@{}}
\toprule
Device & $\xi$ (nm) & $A_\perp$ (nm$^2$) & $A_\parallel$ (nm$^2$) & $t_\mathrm{s}/d^*$ \\ \midrule
1      & 66         & 18000              & 7100              & 0.131              \\
2      & 81         & 23080              & 5718              & 0.1                \\ \bottomrule
\end{tabular}
\end{table}

\subsection*{Asymmetric Little-Parks effect in the non-destructive and destructive regimes}

In this section, we present experimental evidence for strong asymmetries of the Little-Parks oscillations of $I_\mathrm{c}$. The asymmetries emerge when, in addition to $B^\theta_\parallel$, the parallel magnetic field misaligned by an angle $\theta$, we apply $B^\theta_\perp$, a small perpendicular magnetic field misaligned by the same angle $\theta$ (refer to the sketch of the setup in Fig.~\ref{device}d). As a result, the total magnetic field vector has different orientation for positive and negative values, which naturally creates a non-symmetric result in the $B$ axis. We first study the non-destructive regime (Device 1) by comparing Little-Parks measurements for two different $B^\theta_\perp$ values and varying the misalignment angle $\theta$. Secondly, we investigate the asymmetry effect in the destructive regime (Device 2) by instead increasing $B^\theta_\perp$ for a fixed misalignment angle $\theta$.

Figure \ref{fig2}b shows the effect on the LP data of applying $B^\theta_\perp=15$ mT on the sample. In contrast to Fig.~\ref{fig2}a, which shows approximately symmetric LP $I_\mathrm{c}$ oscillations measured at $B^\theta_\perp=0$ mT, the data in Fig.~\ref{fig2}b shows strong $\pm B^\theta_\parallel$ asymmetries in the LP oscillations. The lobe at $-3\Phi_0$ present in the symmetric case of Fig.~\ref{fig2}a  is missing in the asymmetric case in Fig.~\ref{fig2}b, whereas the lobe at $+3\Phi_0$ in Fig.~\ref{fig2}b is larger than the corresponding lobe in Fig.~\ref{fig2}a. As shown in Fig.~\ref{fig2}c, if the direction of $B^\theta_\perp$ is reversed, the LP asymmetries are mirrored along the vertical axis.

A decrease in the misalignment angle $\theta$ has two important consequences: 1) The size of the last lobe increases, due to a smaller effective perpendicular field. This is evidenced in the comparison of Fig.~\ref{fig2}d for $\theta=1.54^\circ$, which shows larger lobes at $\pm 3\Phi_0$ than Fig.~\ref{fig2}a, for $\theta=4.4^\circ$. 2) The degree of asymmetry decreases. To put this in evidence, Fig.~\ref{fig2}e for $\theta=1.54^\circ$ and $B^\theta_\perp=15$ mT can be compared with Fig.~\ref{fig2}b for $\theta=4.4^\circ$ and the same $B^\theta_\perp$ value. The missing lobe at $-3\Phi_0$ in Fig.~\ref{fig2}b reappears in Fig.~\ref{fig2}e for smaller misalignment.     

Our single-cylinder model fully accounts for the observed asymmetries, with $\theta$ as the only parameter which is varied; $\theta=4.4^\circ$ in Figs.~\ref{fig2}a-c and $\theta=1.54^\circ$ in Figs.~\ref{fig2}d,e. Other parameters are the same as those used to fit the data in Fig.~\ref{fig2}a, given above.

 Out of five measured devices, two were found to be in the destructive regime (see SM). Here, we investigate asymmetric Little-Parks oscillations in Device 2, which lies in this regime. The observed phenomenology is similar to that in the non-destructive regime, aside from full destruction of superconductivity at half-flux quanta ($I_\mathrm{c}=0$ at $n\Phi_0/2$). 
 
\begin{figure}[ht!]
    \centering
    \includegraphics[width=1\textwidth]{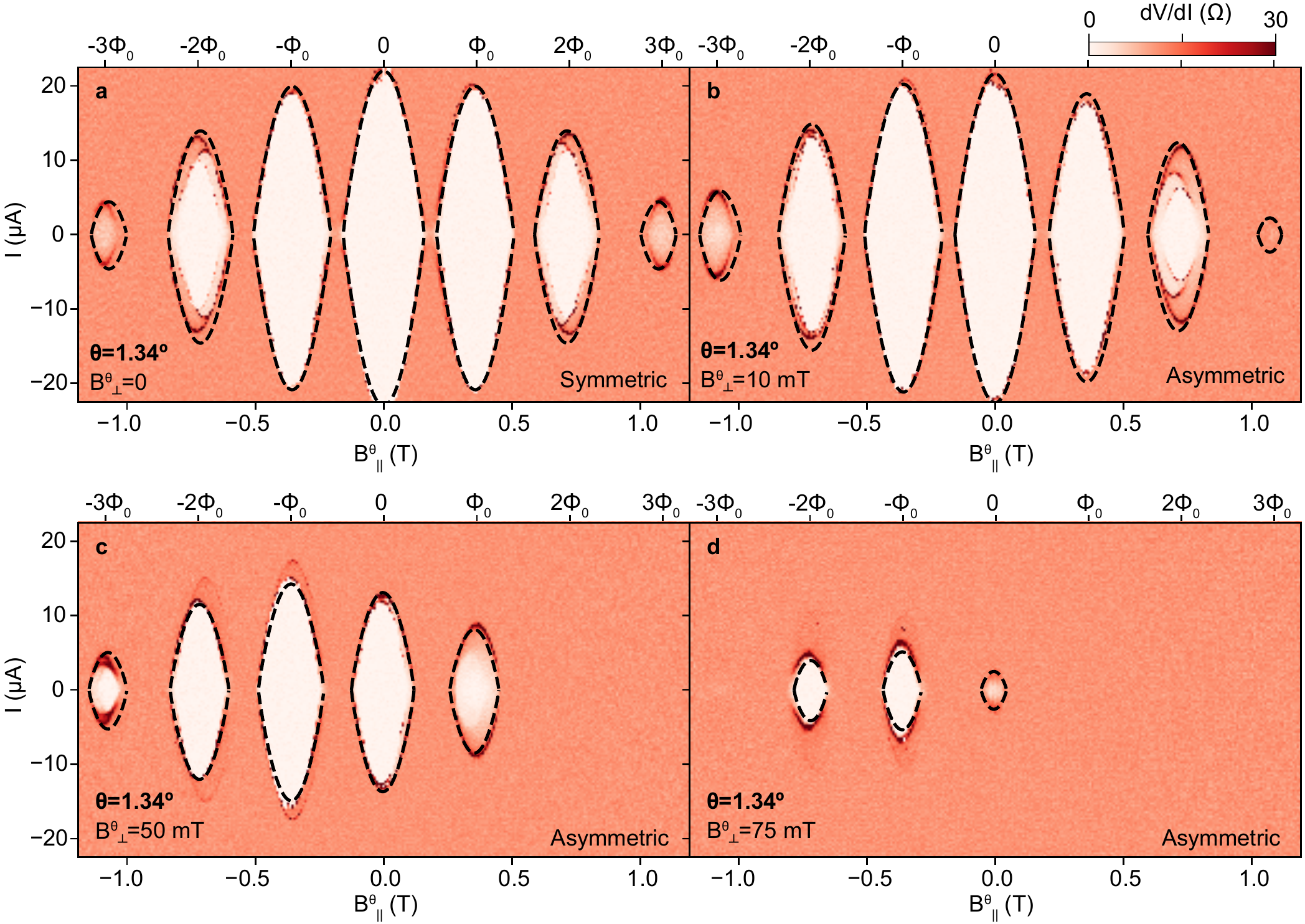}
    \caption{\textbf{Asymmetric Little-Parks effect in the destructive regime (Device 2).}  \textbf{(a-d)} Little-Parks data and fit for Device 2, akin to those shown in Figs.~\ref{fig2}a-c. The misalignment angle $\theta=1.34^\circ$ is kept fixed in (a-d). The misaligned perpendicular field, $B^\theta_\perp$, is progressively increased as indicated in each panel, leading to an increased asymmetry in the Little-Parks oscillations of the critical current, $I_\mathrm{c}$. Fit parameters are given in Tab.~\ref{Tab1}.}
    \label{fig3}
\end{figure}

 Figs.~\ref{fig3}a-d show the evolution of the measured LP oscillations in this device with increasing $B^\theta_\perp$, for fixed $\theta$. In Fig.~\ref{fig3}a, at $B^\theta_\perp=0$, the oscillations are approximately symmetric in $\pm B^\theta_\parallel$. In Fig.~\ref{fig3}b, at $B^\theta_\perp=10$ mT, the lobes at negative $B^\theta_\parallel$ are significantly more pronounced than those at positive $B^\theta_\parallel$. The asymmetry increases significantly in Fig.~\ref{fig3}c at $B^\theta_\perp=50$ mT, with the $2\Phi_0$ and $3\Phi_0$ lobes absent at positive $B^\theta_\parallel$, and the lobe at $-\Phi_0$ becoming larger than the zeroth lobe. In Fig.~\ref{fig3}d, at the largest $B^\theta_\perp$ shown, $B^\theta_\perp=75$ mT, all positive $B^\theta_\parallel$ are absent and the zeroth lobe turns faint in comparison to the $-\Phi_0$ and $-2\Phi_0$ lobes. 
 
 Our model of $I_\mathrm{c}$, shown as dashed lines in Figs.~\ref{fig3}a-d, matches reasonably well the behavior of the lobe boundaries as $B^\theta_\perp$ is increased with a single set of fitting parameters, shown in Tab.~\ref{Tab1}.

\section*{Discussion}

We reported the Little-Parks effect in a new hybrid superconducting platform, consisting of double semiconductor nanowires coated by a superconducting shell. While the semiconductor nanowires were used here only as a template to shape the shell, they can in principle be used in future experiments to explore topological superconductivity in setups involving two hybrid Rashba nanowires~\cite{Beri2012Oct,Klinovaja2014Jul}, using the recent findings involving the LP effect in single hybrid Rashba nanowires coated by a superconducting shell as a starting point~\cite{Vaitiekenas2020Mar}. The hybrid Rashba cores could also be used to extend investigations of Yu-Shiba-Rusinov states in quantum dots coupled to single core-shell nanowires~\cite{Valentini2020Aug}.

We found that, despite their double-nanowire template, the superconducting shell behaved as a \textit{single} hollow superconducting cylinder. Both the destructive and non-destructive LP regimes were observed, indicating a smaller superconducting coherence length in the latter case, and variations in the diameter of the nanowires. In the presence of a small misalignment of the applied parallel and perpendicular magnetic fields with respect to the nominally aligned parallel and perpendicular directions to the axis of the nanowires, the LP oscillations showed strong \textit{asymmetries} in the parallel field direction with respect to zero field. These strong asymmetries may be used to calibrate the alignment of the field with the axis of the nanowires, so as to maximize the critical field of the superconductor and thus maximize the observed number of LP oscillations. Given that a single cylinder model is used to model these asymmetries, the asymmetries are also expected to be present in \textit{single} nanowires coated by a superconducting shell.

To convert parallel and perpendicular magnetic fields to magnetic fluxes, our model uses different areas, $A_\parallel$ and $A_\perp$ \cite{PhysRevB.76.094511}, with $A_\parallel<A_\perp$, in contrast to a previous study in single nanowires coated by superconducting shells, in which $A_\parallel=A_\perp$~\cite{vaitiekenas2020anomalous}. While $A_\parallel$ is interpreted as the cross section of the two nanowires, the physical meaning of $A_\perp$ is not presently understood beyond the phenomenological requirement that $A_\parallel<A_\perp$ to explain the lower perpendicular critical field of the samples (See SM). We note that the model is expected to deviate from the data for field perpendicular to the axis of the nanowires due to the hexagonal cross section of the two nanowires serving as template for the shell, which is different from a strictly circular cross section. The deviation is less important for parallel field, as in this case the field is aligned to the facets of the shell. While $I_\mathrm{c}$ data for field perpendicular to the nanowires (at zero parallel field) is well fitted by the model, $I_\mathrm{c}$ data from rotations of the field at fixed field magnitude is generally not (see SM). This may reflect the expected discrepancy with the model due to the shell geometry, as well as the need for more complex modelling with a realistic geometry of the shell.

The data in Figs.~\ref{fig2} and \ref{fig3} shows additional switching currents at currents above the first switching identified as the critical current. A clear example of additional switchings is seen in Fig.~\ref{fig3}a in the $-2\Phi_0$ lobe. The additional switching currents form a series of higher lobes, which are shifted leftwards or rightwards with respect to the main LP lobes, given by the first switching. The origin of these lobes, which have been previously observed in single nanowires coated by superconducting shells~\cite{vaitiekenas2020anomalous}, is beyond the single cylinder model and it is out of the scope of the present work.

The clamping of the upper segments of these nanowires, which appears to be responsible for the observed single shell behavior, may be avoided by the growth of thicker, less flexible double nanowires~\cite{Kannepreprint2021}. Independent Little-Parks oscillations in the two nanowires may aid in attaining independent pairs of flux-induced Majorana zero modes in each nanowire, while the shared phase winding demonstrated in this work may be of utility to further characterize Majorana zero modes.

\section*{Methods}

\subsection*{Extraction of model parameters}

Here we describe the obtention of the parameters given in Tab.~\ref{Tab1}, used to fit the data from Devices 1 and 2 in Figs.~\ref{fig2} and \ref{fig3} with the single hollow superconducting cylinder model. As the template for the superconducting shell in our devices consists of two nanowires of hexagonal cross section, we converted geometric device parameters into effective single cylinder parameters. The diameter of each nanowire was estimated from the transmission electron micrograph in Fig.~\ref{device}b at $d \approx 90$ nm (including the Al shell). The area $A_\parallel$ of two hexagons of this diameter equals the area of a circle with a diameter $d^*\approx 130$ nm. Additionally, from the same electron micrograph, we obtained $t_\mathrm{s}=13$ nm. The parameters $\theta$, $\xi$ and $A_\perp$ were kept free. Two distinct sets of values for these parameters were found by fitting the corresponding multiple sets of data for Devices 1 and 2 in Figs.~\ref{fig2} and \ref{fig3}. The experimentally measured values of $A_\parallel$, $d^*$ and $t_\mathrm{s}$ were further fine-tuned for a good fit to the data. 


\bibliography{sample}

\section*{Acknowledgements}
We would like to acknowledge fruitful discussions with Michele Burrello, Ida Nielsen, Saulius Vaitiekėnas, Jens Paaske, Gorm Steffensen, Elsa Prada, Pablo San-Rose, Fernando Pe\~{n}aranda and Ramon Aguado.
The project received funding from the European Union’s Horizon 2020 research and innovation program under the Marie Sklodowska-Curie Grant Agreement No. 832645. We additionally acknowledge financial support from the Carlsberg Foundation, the Independent Research Fund Denmark, QuantERA “SuperTop” (NN 127900), European Union’s Horizon 2020 research and innovation programme FETOpen Grant No. 828948 (AndQC), the Danish National Research Foundation, Villum Foundation
Project No. 25310, and the Sino-Danish Center.

\section*{Author contributions statement}

AV fabricated the devices. AV, JCES and SL performed the measurements. AV, JB, JCES, JN and KGR analyzed the data. TK, MM, DO and JN developed and characterized the double nanowires. AV and JCES wrote the manuscript with comments from all authors.

\section*{Additional information}

Supplementary Materials are available.

\end{document}